**Spontaneous segregation of visual information between parallel streams of a multi-stream convolutional neural network**

**Hiroshi Tamura**[1,2,*]

[1]Graduate School of Frontier Biosciences, Osaka University, Suita, Osaka 565-0871, Japan

[2]Center for Information and Neural Networks, Suita, Osaka 565-0871, Japan

**Number of pages**: 37

**Number of figures**: 8

**Number of tables**: 2

**Word counts**: Abstract, 142; Introduction, 556; Discussion, 711



**Abstract**

Visual information is processed in hierarchically organized parallel pathways in the primate brain. In lower cortical areas, color information and shape information are processed in a parallel manner, while in higher cortical areas, various types of visual information, such as color, face, animate/inanimate, are processed in a parallel manner. In the present study, the possibility of spontaneous segregation of visual information in parallel streams was examined by constructing a convolutional neural network with parallel architecture in all of the convolutional layers. The results revealed that color information was segregated from shape information in most model instances. Deletion of the color-related stream decreased recognition accuracy in the inanimate category, whereas deletion of the shape-related stream decreased recognition accuracy in the animate category. The results suggest that properties of filters and functions of a stream are spontaneously segregated in parallel streams of neural networks.



**Introduction**

In the cerebral cortex, visual information is processed in hierarchically organized parallel pathways (Livingstone & Hubel, 1984; Livingstone & Hubel, 1988; Felleman & Van Essen, 1991; DeYoe et al., 1994; Sincich & Horton, 2005; Nassi & Callaway, 2009; Kandel et al., 2021). The dorsal pathway processes spatial and motion information, whereas the ventral pathway processes object information. Even within the ventral pathway, information is processed in a parallel and independent manner. For example, in the lower visual cortical areas, such as the primary visual cortex (V1) and secondary visual area (V2), color information and orientation information are processed in different cortical modules (Livingstone & Hubel, 1988; Ts'o & Gilbert, 1988; Peterhans & von der Heydt, 1993; Levitt et al., 1994; Leventhal et al., 1995; Gegenfurtner et al., 1996; Tamura et al., 1996; Landisman & Ts'o, 2002; Shipp & Zeki, 2002; Nassi & Callaway, 2009; Economides et al., 2011; Garg et al., 2019; Peres et al., 2019). Furthermore, lower spatial frequency (SF)-preferring neurons are found in a specific compartment, while higher SF-preferring neurons are found in another compartment in V1 (Silverman et al., 1988; Tootell et al., 1988). In the higher visual cortical areas, color information and shape information are processed in a segregated manner (Komatsu et al., 1992; Tamura & Tanaka, 2001; Tanigawa et al., 2010; Lafer-Sousa & Conway, 2013). In addition to the segregation of color and shape, animate images are processed in a segregated manner from inanimate images (Caramazza & Shelton, 1998; Kriegeskorte et al., 2008; Naselaris et al., 2012; Bao et al., 2020).

The present study examined the ways in which information is segregated in a parallelized



convolutional neural network (CNN) to gain insight into the origin of functional segregation of visual information in the ventral stream. Specifically, the possibility of spontaneous segregation of visual information in parallel streams was examined by comparing filter properties and examining the effects of deletion of a stream. CNNs for visual object recognition have been constructed on the basis of the architecture and functions of visual cortical areas (Fukushima, 1980; Krizhevsky et al., 2012), and consist of hierarchically organized layers that have many filters. Outputs from filters in the lower layer of CNNs show similarity to the lower areas of the primate visual cortices and those in higher layers show similarities to higher cortical areas (Khaligh-Razavi & Kriegeskorte, 2014; Yamins et al., 2014; Güçlü & van Gerven, 2015; Yamins & DiCarlo, 2016; Flachot & Gegenfurtner, 2018; Wagatsuma et al., 2022). Thus, CNNs for visual object recognition provide an ideal model for the primate ventral visual pathway. Furthermore, analysis using CNNs for visual object recognition provides a unique opportunity to understand the functioning of neurons in visual cortical areas (Kanda et al., 2020; Leavitt & Morcos, 2020; Dobs et al., 2022; Kanwisher et al., 2023).

In the current study, I constructed a modified version of AlexNet (Krizhevsky et al., 2012), called two-streams fully parallel (2SFP) AlexNet (Fig. 1A). Previous CNN studies introduced parallel architecture, in conv1 and conv2 (Krizhevsky et al., 2012; Flachot & Gegenfurtner, 2018), or using architecture different from the AlexNet (Feichtenhofer et al., 2019; Bakhtiari et al., 2021; Nayebi et al., 2021). I introduced parallel architecture to AlexNet in all convolutional layers. This architecture allows comparison of filter properties both in lower and higher layers, and analysis of the effect of deletion of a stream.



**Results**

The present study was based on 16 instances (Table 1) of 2SFP-AlexNet with some variation in the initial learning rate and batch size. In addition, one instance of 2SFP-VGG11 and one instance of three-streams fully parallelized (3SFP) AlexNet were also examined (Table 1).

**\*\*\*\*\* Table 1 near here \*\*\*\*\***

**Properties of filters in convolutional layer 1 of two-streams fully parallelized AlexNet**

After training, conv1 filters acquired a variety of kernels. Some filters were color selective while others were orientation selective (Fig. 1B), and some filters preferred lower spatial frequency while others preferred higher spatial frequency. A conv1 filter (rightmost filter in the second row of Fig. 1B-left, \*) of stream 1 of a model instance preferred red color, showed no orientation selectivity and preferred lower spatial frequency. Color index and orientation index of the filter were 0.994 and 0.0273, respectively, and preferred spatial frequency (SF) was 0 (i.e., direct current [DC]). A conv1 filter (second filter in the top row of Fig. 1B-left, \*\*) of stream 2 of the same model instance showed no color selectivity (color index, 0.00102) but preferred an oblique orientation (orientation index, 0.770) and preferred a middle spatial frequency (preferred SF, 2 cycles/filter).

**\*\*\*\*\* Figure 1 near here \*\*\*\*\***



Degree of color selectivity, degree of orientation selectivity, and preferred SF were related to each other. In the instance shown in Figure 1B-right, the color index was negatively correlated with the orientation index ($r = -0.57$, n = 117 [11 filters with flat kernel were excluded from the analysis], Spearman's rank correlation; Fig. 2A-left) and with preferred SF ($r = -0.66$; Fig. 2A-center), and the orientation index was positively correlated with preferred SF ($r = 0.68$; Fig. 2A-right). These relationships were consistently observed in all 16 instances (Fig. 2B, C). These results suggested that color information and orientation information were encoded by different populations of filters, and color-selective filters were less orientation selective and tended to prefer lower SF, while orientation-selective filters were less color selective and preferred higher SF. Although the color index was negatively correlated with the orientation index, there was a small but significant fraction of filters that simultaneously had a higher color index as well as a higher orientation index (Fig. 2 A and C, left panel, points around the upper right corner; see for example, the 42nd filter, 2nd filter in the 6th row of stream 2 of Fig. 1B-right, horizontally oriented yellow and blue filter), suggesting that some filters were selective to both color and orientation (Garg et al., 2019).

**\*\*\*\*\* Figure 2 near here \*\*\*\*\***

The results described above raise the question of how these conv1 filters are associated with the two streams of 2SFP-AlexNet. I found that color-selective filters were numerous in a stream, and orientation-selective filters were numerous in the other stream in most instances. As a result, selectivity indices and preferred SF of conv1 filters differed between the two streams of



2SFP-AlexNet. For example, the median color index values of stream 1 and 2 of Figure 1B-left were 0.46 and 0.0050, respectively, and the median orientation index values of stream 1 and 2 were 0.33 and 0.77, respectively. These indices differed between streams (color index, $p = 4.93 \times 10^{-12}$; orientation index, $p = 3.58 \times 10^{-6}$; Mann–Whitney U test; Fig. 1C-left). Preferred spatial frequency also differed between the two streams ($p = 9.05 \times 10^{-10}$; Fig. 1C-left). The mean preferred SFs of stream 1 and 2 were 0.56 and 1.91, respectively. As a result, in the instance shown in Figure 1B-left, conv1 in stream 1 had filters with a higher degree of color selectivity, a lower degree of orientation selectivity, and lower preferred SF than those in the other stream.

Significant differences in color index values, orientation index values, and preferred SF were also observed in the instance shown in Figure 1B-right. In this instance, conv1 in stream 1 had filters with a lower degree of color selectivity, a higher degree of orientation selectivity and preferred higher SF compared with those in the other stream (Fig. 1C-right). Among the 16 instances of 2SFP-AlexNet, differences in the color index and orientation index were observed in 12 and 10 instances, respectively (Fig. 3A). Differences in preferred SF were observed in 10 instances (Fig. 3A). Differences in color index values, orientation index values, and preferred SF were simultaneously observed in eight instances (Fig. 3A). In all eight instances, a stream tended to have conv1 filters with strong color selectivity, weak orientation selectivity, and a preference for lower SF, and the other stream tended to have conv1 filters with weak color selectivity, strong orientation selectivity, and a preference for higher SF. In Figure 3B, the median color index, median orientation index, and mean preferred SF of conv1 filters of



stream1 were plotted against those of stream2. In general, if a median index of a stream was high, the median index of the other stream was low, and there was a negative correlation between the index (−0.67 to −0.81; Fig. 3B), also suggesting the segregation of filter properties between streams.

However, such segregation was not observed in all model instances. For example, in the instance shown in Figure 1B-center, the median color index values of conv1 filters of stream 1 and 2 were 0.0068 and 0.019, respectively, and the index did not differ between the streams ($p = 0.188$; Fig. 1C-center). The median orientation index of conv1 filters of stream 1 and 2 were 0.73 and 0.51, respectively, and the index did not differ between the streams ($p = 0.030$; Fig. 1C-center). The mean preferred SF of conv1 filters of stream 1 and 2 were 1.77 and 1.44, respectively, and preferred SF also did not differ between the streams ($p = 0.073$; Fig. 1C-center). Furthermore, even in the instance with a significant difference in indices, the degree of difference in indices and preferred SF varied across instances. The plots shown in Figure 3B revealed that some points were closer to the equality line while others were further away from it, suggesting that a degree of segregation varied among instances. It could be speculated that such a degree of segregation is related to hyperparameters of AlexNet (see Table 1). Indeed, there was a tendency for small batch size to cause a higher degree of segregation. However, even among instances with the same batch size, there were substantial variations in the degree of segregation.

**\*\*\*\*\* Figure 3 near here \*\*\*\*\***



**Properties of filters in convolutional layer 2–5 of two-streams fully parallelized AlexNet**

To examine the properties of filters in higher convolutional layers, the stimulus image that induced large activation in each filter (most effective stimulus [MES]) was calculated (Fig. 4A, B). MESs of some filters of conv2–5 were colorful, whereas some others were colorless (see Fig. 4A, B). Higher SF component was stronger in some MESs, whereas lower SF component was stronger in some other MESs. Degree of color selectivity and preferred SF of MESs were related to each other in conv2–5, and color selective MESs tended to contain a lower SF component, whereas color-non-selective MESs contained a variety of SFs. As a result, color index values were negatively correlated with preferred SF in conv2–5 ($r = -0.44$ to $-0.35$, Spearman's rank correlation; Fig. 5A).

Filter properties of conv2–5 examined using MES differed between streams. In the instance shown in Figure 4A, the color index of MESs of conv2–5 of stream 1 (0.68–0.76, median) was larger than that of stream 2 (0.14–0.19; $p = 3.78 \times 10^{-56}$–$6.10 \times 10^{-127}$, Mann-Whitney U test; Fig. 4C-left). Preferred SF of MESs also differed between the two streams ($p = 1.87 \times 10^{-14}$–$1.21 \times 10^{-80}$; Fig. 4C-right) and preferred SF of conv2–5 of stream 1 (2–4, median) was lower than that of stream 2 (14–15, median). Significant differences in color index values of MESs of conv2, 3, 4, and 5 were observed in 15, 15, 16, and 16 instances, respectively (Fig. 5B). A significant difference in preferred SF of MESs of conv2, 3, 4, and 5 was observed in 14, 16, 15, and 16 instances, respectively. The median color index value of a stream was negatively correlated with that of the other stream ($-0.78$ to $-0.89$; Fig. 5B), and the median preferred SF



of a stream was also negatively correlated with that of the other stream (−0.65 to −0.84; Fig.

5C). Thus, color-selective filters that preferred lower SF in conv2–5 were segregated from

color-non-selective filters that preferred higher SF in conv2–5. The plots also revealed that

some points were closer to the equality line while others were further away from it, suggesting

that the degree of segregation varied among instances.

In the instance shown in Figure 4A, MESs of filters of conv2–5 in stream 1 appear colorful,

while those in stream 2 appear colorless. These properties may be derived from the properties of

conv1 filters, because color-selective filters were concentrated in conv1 of stream 1 and

colorless filters were concentrated in conv1 of stream 2 of the instance (see Fig. 1B-left).

However, in another instance of Figure 4B, where color-selective filters were observed in both

conv1 of stream 1 and stream 2 (see Fig. 1B-center) and color index values did not differ

between streams, MESs of filters of conv2–5 in stream 2 appeared more colorful than those of

stream 1. To examine whether the difference in color selectivity and SF preference of conv2–5

was inherited from those of conv1, the correlation between the absolute difference in color

index of conv1 filters and that of conv2–5 was examined. There was a positive correlation

between these measures in conv2–5 ($r = 0.61$–$0.80$, Spearman's rank correlation; Fig 5D-left),

suggesting that the difference in color index of MESs of filters of conv2–5 is likely to be

inherited from the difference observed in conv1. In contrast, the correlation between the

absolute difference in preferred SF of conv1 filters and that of conv2–5 was weak ($r = 0.21$–

$0.41$, Spearman's rank correlation; Fig 5D-right). These results suggest that the difference in

color selectivity was inherited from that of conv1, while this tendency was weak for SF



preference.

**\*\*\*\*\* Figure 4 near here \*\*\*\*\***

**\*\*\*\*\* Figure 5 near here \*\*\*\*\***

**A comparison of stimulus representation between two-streams fully parallelized AlexNet**

To examine how the difference in filter properties contributes to the difference in information

representation between streams, I compared the representation of a set of 1,000 stimulus images

between streams of 2SFP-AlexNet by calculating representational dissimilarity matrix (RDM;

Kriegeskorte et al., 2008; Fig. 6A). RDM of conv1 of stream1 was similar to that of stream 2

(Fig. 6A-left) despite the difference in the degree of color and orientation selectivity and

preferred SF (see Fig. 1B-left). Similarity in stimulus representation was quantified by

calculating the correlation coefficient between the RDMs (Fig. 6B). In the instance of Figure

6A, the correlation coefficient of RDM of conv1 between the two streams was 0.80 (Fig. 6B).

Note that large differences in color index, orientation index, and SF preference of conv1 filters

between streams were observed in the instance of Figure 6A, B (see Fig. 1B-left). This result

suggests that similarity in image representation of conv1 filters between streams was not related

to similarity in degree of color and orientation selectivity and SF preference. Indeed, the

correlation coefficients of RDM of conv1 filters between streams obtained with all 16 instances

were always high (0.71−0.95) and were not related to the absolute difference in color index

between streams ($r = 0.13$, Spearman's rank correlation; Fig. 6C).





Contrary to RDM of the conv1, RDM of conv5 of stream1 was different from that of stream 2 (Fig. 6A-right). In the instance of Figure 6A, the correlation coefficient of RDM of conv5 between the two streams was 0.31 (Fig. 6B). Thus, the correlation coefficients between RDMs from different streams at the same hierarchical level decreased gradually along the hierarchy of 2SFP-AlexNet. This tendency was confirmed for all 16 instances (Fig. 6D). The correlation coefficient of RDMs differed among conv layers ($p = 1.40 \times 10^{-11}$, Friedman test for repeated samples), and the correlation coefficient of RDMs of conv5 (0.33, median) was smaller than that of conv1 (0.82). The result suggests that similarity in information representations between two streams decreased during hierarchical processing from conv1 to conv5 and representations became less correlated between streams.

### Effects of deletion of a stream of two-streams fully parallelized AlexNet on the classification of images

If each of the two streams of 2SFP-AlexNet represents images in a different manner, the effect of deleting one stream on classification accuracy is likely to be different from that of deleting the other stream. Indeed, deletion of stream 1 of the instance in Figure 7 resulted in the largest decrease in the proportion of correct responses in the "rapeseed" category (Fig. 7-left, filled blue circle), whereas deletion of stream 2 resulted in the largest decrease in the proportion of correct responses in the "zebra" category (Fig. 7-right, filled orange circle). Note that deletion



of stream 1 did not affect the proportion of correct responses in the "zebra" category (Fig. 7-left, filled orange circle) and deletion of stream 2 did not affected the proportion of correct responses in the "rapeseed" category (Fig. 7-right, filled blue circle), demonstrating double dissociation. This result suggests that deletion of a stream of 2SFP-AlexNet affected classification of an image category in a specific manner.

***** **Figure 7 near here** *****

Among the 32 streams from 16 instances of 2SFP-AlexNet, deletion of 13 streams resulted in the largest decrease in proportion of correct responses in the "zebra" category, and that of the other 13 streams resulted in the largest decrease in the proportion of correct responses in the "rapeseed" category (Table 2). In the remaining cases, deletion resulted in the largest decrease in the proportion of correct responses in "giant panda," "porcupine," "dugong," "European fire salamander," "maypole," and "ambulance" categories. Because "zebra," "giant panda," "porcupine," "dugong," and "European fire salamander" are animate objects, and "rapeseed," "maypole," and "ambulance" are inanimate objects, these categories can be divided into animate and inanimate categories. The largest decrease in the proportion of correct responses in the inanimate category after deletion of a stream was observed in 15 streams. Similarly, the largest decrease in proportion correct in the animate category after deletion of a stream was observed in 17 streams (Table 2). Importantly, if the deletion of a stream resulted in the largest decrease in the inanimate category, deletion of the other stream resulted in the largest decrease in the animate category in 15 among the 16 instances. These results suggest that the animate and



inanimate categories are represented by different streams.

**\*\*\*\*\* Table 2 near here \*\*\*\*\***

Filter properties were compared between the inanimate and animate streams. If deletion of a stream resulted in the largest decrease in the inanimate category, the stream was designated as an inanimate stream, whereas if deletion of a stream resulted in the largest decrease in the animate category, the stream was designated as an animate stream. Color index of conv1 filters differed between the inanimate and animate streams ($p = 1.37 \times 10^{-4}$, Mann-Whitney U test), and the inanimate stream had a higher color index (0.13, median) compared with the animate stream (0.0068). Orientation index and preferred SF also differed between the inanimate and animate streams (orientation index, $p \approx 0$; SF preference, $p = 1.16 \times 10^{-3}$). Orientation index of the inanimate stream (0.40, median) was lower than that of the animate stream (0.75; Fig. 7B), and preferred SF of the inanimate stream (1.36, mean) was lower than that of animate stream (1.77). Conv2−5 filters of the inanimate stream also had higher color index and lower preferred SF than those of animate stream (color index, $p = 5.86 \times 10^{-6}$−$8.35 \times 10^{-6}$; preferred SF, $p = 2.12 \times 10^{-6}$−$4.50 \times 10^{-5}$; Fig. 7C). These results suggest that the inanimate stream consists of color selective and weakly orientation selective and lower SF-preferring filters, whereas the animate stream consists of weakly color selective and orientation selective and higher SF-preferring filters.

**Properties of two-streams fully parallelized VGG11 and three-streams fully parallelized**



**AlexNet**

To examine whether the segregation of functional properties between two streams of 2SFP-AlexNet was observed in another type of convolutional neural network, VGG11 (Simonyan & Zisserman, 2015) was parallelized to construct 2SFP-VGG11. 2SFP-VGG11 has two streams of eight hierarchically organized convolutional layers and five pooling layers. Outputs from each stream were combined and fed into fully connected layers, then to the output layer. 2SFP-VGG11 was randomly initialized and trained for classification of 1,000 object categories using the ImageNet database (Deng et al., 2009). Similar to 2SFP-AlexNet, segregation of filters according to their properties was observed in 2SFP-VGG11 (Fig. 8A). Color index of conv1 of stream 1 (0.00082, median) was smaller than that of stream 2 (0.022, median; $p = 0.0096$, Mann–Whitney U test). Orientation selectivity and preferred SF of conv1 filters were not examined, because of the small size ($3 \times 3$) of conv1 filters of the 2SFP-VGG11.

Filter properties of conv2–8, which was examined using MES, also differed between streams of 2SFP-VGG11. The color index of conv2–8 of stream 2 (0.43–0.71, median) was larger than that of stream 1 (0.021–0.042; $p = 7.14 \times 10^{-169}$–$6.24 \times 10^{-41}$, Mann–Whitney U test). Preferred SF also differed between the two streams ($p = 6.36 \times 10^{-70}$–$1.75 \times 10^{-25}$) and preferred SF of conv2–8 of stream 2 (1–39, median) was lower than that of stream 1 (10–62, median). Thus, color-selective filters that preferred lower SF in conv2–8 were segregated from color-non-selective filters that preferred higher SF in conv2–8 of the instance of the 2SFP-VGG11.

The largest decrease in the proportion of correct responses in the animate category ("zebra")



after deletion of stream 1 of 2SFP-VGG11 was observed, and the largest decrease in the

proportion of correct responses in the inanimate category ("rapeseed") was observed after

deletion of stream 2. Stream 1 was less color selective than the other stream and the most

affected image category after deletion of the stream was animate, whereas stream 2 was more

color selective than the other stream and the most affected image category after deletion of the

stream was inanimate. The result was similar to that obtained with 2SFP-AlexNet; the inanimate

stream consisted of color-selective filters, whereas the animate stream consisted of weakly

color-selective filters. Thus, although the architecture of networks of 2SFP-VGG11 differed

from that of 2SFP-AlexNet, properties of parallel streams of 2SFP-VGG11 were similar to those

of 2SFP-AlexNet.

**\*\*\*\*\* Figure 8 near here \*\*\*\*\***

Segregation of functional properties across 3SFP-AlexNet was also examined. Color index,

orientation index, and preferred SF of conv1 filters differed among three streams ($p = 2.99 \times 10^{-16}$–$7.67 \times 10^{-11}$, Kruskal–Wallis H-test; Fig. 8B). Conv1 filters of stream 1 were mostly

orientation selective (0.61, median orientation index), but color selectivity was low (0.018,

median color index), and preferred modest SF (0.90, mean preferred SF). Conv1 filters of

stream 2 were mostly color selective (0.66, median color index), but weakly selective to

orientation (0.25, median orientation index) and preferred lower SF (0.48, mean preferred SF).

Conv1 filters of stream 3 were also mostly orientation selective (0.85, median orientation

index), but color selectivity was low (0.0021, median color index), and higher SF was preferred



(2.31, mean preferred SF).

Filters in higher convolutional layers of 3SFP-AlexNet exhibited a similar tendency. The color index of MES of conv2–5 of stream 2 (0.72–0.79, median) was larger than that of streams 1 (0.26–0.38) and 3 (0.043–0.077; $p = 8.58 \times 10^{-216}$–$2.37 \times 10^{-84}$, Kruskal–Wallis H-test). Preferred SF of MES also differed among the three streams ($p = 8.92 \times 10^{-119}$–$7.48 \times 10^{-38}$) and preferred SF of conv2–5 of stream 2 (1–4, median) was lower than that of streams 1 (5–12, median) and 3 (16–35, median). Thus, if there are three streams, a stream contains color-selective and low SF-preferring filters, another stream contains orientation-selective and high SF-preferring filters, and yet another stream contains orientation-selective and modest SF-preferring filters. Similar segregation in multiple streams of parallelized or branched CNNs has been reported previously (Voss et al., 2021).

The largest decrease in the proportion of correct responses in the inanimate category ("rapeseed") after deletion of stream 2 was observed, and the largest decrease in proportion correct in the animate category ("West Highland white terrier" and "porcupine") was observed after deletion of streams 1 or 3 of 3SFP-AlexNet. Thus, stream 2, which has many color-selective filters, was involved in the classification of the inanimate category, while streams 1 and 3, which have many orientation-selective filters, were involved in classification of the animate category.

**Discussion**



The main finding of the present study is that color/inanimate information are segregated from shape/animate information in parallel streams of the CNN (Fig. 9). The results suggest that properties of filters and functions of a stream are spontaneously segregated in parallel streams of the CNN without intentionally assigning a particular property and function to a stream.

In the present study, I constructed a modified version of AlexNet (i.e., 2SFP-AlexNet), which has two fully parallelized streams from conv1 to conv5. Introduction of parallel architecture throughout the convolutional layers allowed analysis of information segregation in lower as well as in higher convolutional layers. Furthermore, analysis of the effects of deletion of a stream becomes possible with this architecture. CNNs with parallel streams have been constructed in previous studies (Krizhevsky et al., 2012; Flachot & Gegenfurtner, 2018; Feichtenhofer et al., 2019; Bakhtiari et al., 2021; Nayebi et al., 2021). The original AlexNet introduced parallel architecture in conv1 and conv2 (Krizhevsky et al., 2012). Similar to the original AlexNet, filters of 2SFP-AlexNet acquired a variety of kernels, and color-selective filters were less orientation selective and tended to prefer lower SF, while orientation-selective filters were less color selective and preferred higher SF. These properties are consistent with the properties of neurons in V1 (Johnson et al., 2001). In the original AlexNet (Krizhevsky et al., 2012), color-agnostic kernels were spontaneously segregated from color-specific kernels in conv1. The present results also revealed spontaneous segregation of color-selective kernels and color-non-selective and orientation-selective kernels between parallel streams in conv1 for most model instances.



Despite the significant difference between parallel steams in color selectivity, orientation selectivity and SF preference of conv1 filters, RDM analysis revealed that image representation of conv1 filters was similar between streams. The results suggest that similarities in image representation can be independent from similarities in color selectivity, orientation selectivity, and SF preference. In conv1, the two streams of parallel streams shared the same receptive field (RF) size, and the similarity in image representation is likely to be derived from the similarity in RF size of conv1 filters. The results also suggest that image representation in the color compartment is similar to that in the shape compartment in early visual cortical areas of the primate brain. RDM analysis also revealed that similarity in RDMs between two streams decreased along the hierarchy, meaning that the information they encoded became less correlated along the hierarchy. This suggests that hierarchically organized parallel pathways create independent information representation between parallel streams.

Creating independent information representations may be related to the specialization of each stream for animate or inanimate classification. The deletion results revealed that the largest decrease in recognition accuracy in the animate category was observed following deletion of a stream, and that in the inanimate category was observed following deletion of the other stream. In the primate brain, animate images are processed in a segregated manner from inanimate images (Caramazza & Shelton, 1998; Kriegeskorte et al., 2008; Naselaris et al., 2012; Bao et al., 2020). Gradual acquisition of independent information representation between streams might be a consequence of the specialization of streams for animate or inanimate information, or might contribute to the specialization.



The color-shape segregation in parallel streams may be a byproduct to produce independent representation between streams and/or to segregate animate-related information from inanimate-related information between parallel streams. Krizhevsky et al. (2012) found spontaneous segregation of color-agnostic kernels from color-specific kernels between conv1 of parallel streams in every model instance, but Flachot and Gegenfurtner (2018) and the current study found that the degree of segregation of color-selective kernels and color-non-selective kernels varied among model instances. Interestingly, there is large variation in the results of physiological studies that examined segregation of color-selective neurons and orientation-selective neurons in compartments revealed by cytochrome oxidase staining (Livingstone & Hubel, 1988; Ts'o & Gilbert, 1988; Peterhans & von der Heydt, 1993; Levitt et al., 1994; Leventhal et al., 1995; Gegenfurtner et al., 1996; Tamura et al., 1996; Landisman & Ts'o, 2002; Shipp & Zeki, 2002; Economides et al., 2011; Garg et al., 2019; Peres et al., 2019). The low consistency of color-shape segregation across 2SFP-AlexNet instances and across physiological experiments suggests that segregation of color information and shape information in parallel streams of CNNs and the primate brain may not be an inevitable organization.



**Methods**

2SFP-AlexNet was constructed and trained using the PyTorch framework (v.1.12.0; Paszke et al., 2019). 2SFP-AlexNet contains two streams of five hierarchically organized convolutional layers (conv1−5) and three pooling layers (Fig. 1A). Outputs from the two streams were combined and fed into fully connected layers, then to the output layer. 2SFP-AlexNet was initialized randomly and trained for classification of 1000 object categories using the ImageNet database (Deng et al., 2009), which contains 1.2 million training images and 50,000 validation images. The size of images was 224 × 224 pixels. The training was performed using stochastic gradient descent (Kiefer and Wolfwitz, 1952) with cross-entropy loss (Murphy, 2012). The number of epochs was 90. The initial learning rate was 0.01, but was 0.005 or 0.02 in some instances to see the effect of learning rate on the degree of information segregation (Table 1). The learning rate was reduced two times every 30 epochs by 0.1. The momentum was 0.9. The batch size was 128, but 16, 32, or 512 images were tested in some instances to examine the effect of batch size on the degree of information segregation (Table 1). After training, top-5 accuracy was approximately 50% with the validation set. Although the performance was lower than the original AlexNet model, filters were well trained and matured for the present purpose.

Color selectivity and orientation selectivity of each filter of conv1 layer were quantified with selectivity indices. If a filter did not develop any structure (i.e., flat kernel; for example, see the 4th filter in the first row of stream 1 of Fig. 1B-right), the filter was excluded from the analyses of index. Color selectivity was evaluated by calculating the correlation coefficient ($r$) of filter weight among red (R), green (G) and blue (B) channels. If a filter was not color selective,



weight values were correlated among channels. The smallest correlation coefficient among the three correlation coefficients ($r_{min}$) was selected, and color index was obtained with the following formula:

Color index = $r_{min} \times (-0.5) + 0.5$

If kernels of one- or two-color channels were flat, variance was zero and $r$ could not be defined. In this case, however, it is obvious that the filter was color selective and color index was set to one. The color index took a value between zero and one, and the larger the color index, the higher the color selectivity. Orientation selectivity was quantified with the following formula after two-dimensional discrete Fourier transform:

Orientation index = $(\text{Amplitude}_p - \text{Amplitude}_o) / (\text{Amplitude}_p + \text{Amplitude}_o)$

Here, $\text{Amplitude}_p$ and $\text{Amplitude}_o$ are the filter weight amplitude at preferred and orthogonal orientation, respectively. Amplitude was calculated by summating the amplitude within $\pm 15°$ and was examined with an interval of 30°. Orientation index was calculated using the preferred color channel, which has the largest weight amplitude. Orientation index takes a value between zero and one, and larger the orientation index, higher the orientation selectivity. Preferred spatial frequency (SF, cycles/filter) of each filter of conv1 layer was examined by summating amplitude along the circumference at each frequency using the preferred color channel. Because the size of conv1 filters was 11 × 11, SF was examined from zero (DC) to 5 cycles/filter.

To examine the properties of filters in higher convolutional layers (conv2−5), which have more than three channels and filter weights were difficult to visualize with RGB values, the stimulus image (most effective stimulus, MES) that induced large activation in each filter was calculated



using gradient ascent starting from an initial image with random RGB values (Erhan et al., 2009; Olah et al., 2017). The mean across all the units that constitute a filter was maximized. The image size was 224 × 224, which was the same as that of the images used in the training and validation sets. Color selectivity of MES was evaluated by calculating the correlation coefficient ($r$) of RGB values among RGB channels. The smallest correlation coefficient among the three correlation coefficients ($r_{min}$) was selected, and color index values were obtained with the following formula.

Color index = $r_{min}$ × (−0.5) + 0.5

Orientation selectivity was not quantified because many of the MESs did not display clear selectivity to orientation. Preferred spatial frequency (SF, cycles/image) of each MES of conv2−5 was examined by summating amplitude along the circumference at each frequency at the preferred color. Because the size of the filters was 224 × 224, SF was examined from zero (DC) to 112 cycles/image.

To compare representation of a set of stimulus images between streams of 2SFP-AlexNet, RDM (Kriegeskorte et al., 2008) was calculated. From each of 1000 categories of the validation set of ImageNet, one stimulus image was randomly selected and created a set of 1000 stimulus images for RDM analysis. The set was consistently used in the present analysis. Filter outputs were calculated to each stimulus image. For example, in the case of conv1, outputs from 64 × 55 × 55 filters were calculated. Normalized distances between outputs of the set of filters to a pair of stimulus images were then calculated (see Fig. 6A). Once RDM for each convolutional layer was calculated, similarity in stimulus representation was quantified by calculating the



correlation coefficient between the RDMs (see Fig. 6B).

To examine the contribution of each stream to image classification, a deletion experiment was performed. To delete a stream, output values of last max-pool layer of the stream was forced to set to zero during the validation trial. The correct proportion was calculated with the validation set. Changes in accuracy for each category were examined, and the most affected category, which showed the largest decrease in accuracy, was clarified.

*Statistical analysis*

All data were pooled for statistical analyses. Analyses were performed with pandas, numpy, scipy, scikit-learn, and visualized with matplotlib and seaborn on Python. The statistical tests used in the present study were the Mann–Whitney U test (two-tailed), Friedman test for repeated samples, and Kruskal–Wallis H-test. The statistical threshold for $p$-values was set at 0.01. Median values were calculated to represent a population except for the SF of conv1, in which the median could not capture the difference between groups and the mean value was calculated.

**Data availability**

The datasets generated during and/or analyzed during the current study are available from the corresponding author on reasonable request.

**Code availability**



The computer codes used during the current study are available from the corresponding author on reasonable request.




**References**

Bakhtiari S, Mineault P, Lillicrap T, Pack C, Richards B (2021) The functional specialization of visual cortex emerges from training parallel pathways with self-supervised predictive learning. 35th Conference on Neural Information Processing Systems (NeurIPS 2021).

Bao P, She L, McGill M, Tsao DY (2020) A map of object space in primate inferotemporal cortex. Nature 583:103-108.

Caramazza A, Shelton JR (1998) Domain-specific knowledge systems in the brain the animate-inanimate distinction. J Cogn Neurosci 10:1-34.

Deng J, Dong W, Socher R, Li L, Li K, Fei-Fei L (2009) ImageNet: A large-scale hierarchical image database. 2009 IEEE Conference on Computer Vision and Pattern Recognition, Miami, FL, pp. 248-255.

DeYoe EA, Felleman DJ, Van Essen DC, McClendon E (1994) Multiple processing streams in occipitotemporal visual cortex. Nature 371:151-154.

Dobs K, Martinez J, Kell AJE, Kanwisher N (2022) Brain-like functional specialization emerges spontaneously in deep neural networks. Sci Adv 8:eabl8913.

Economides JR, Sincich LC, Adams DL, Horton JC (2011) Orientation tuning of cytochrome oxidase patches in macaque primary visual cortex. Nat Neurosci 14:1574-1580.

Erhan D, Bengio Y, Courville A, Vincent P (2009) Visualizing higher-layer features of a deep network. University of Montreal, 1341:3.

Feichtenhofer C, Fan H, Malik J, He K (2019) SlowFast networks for video recognition. arXiv:1812.03982.

Felleman DJ, Van Essen DC (1991) Distributed hierarchical processing in the primate cerebral





cortex. Cereb Cortex 1:1-47.

Flachot A, Gegenfurtner KR (2018) Processing of chromatic information in a deep convolutional neural network. J Opt Soc Am A Opt Image Sci Vis 35:B334-B346.

Fukushima K (1980) Neocognitron: A self-organizing neural network model for a mechanism of pattern recognition unaffected by shift in position. Biol Cybern 36:193–202.

Garg AK, Li P, Rashid MS, Callaway EM (2019) Color and orientation are jointly coded and spatially organized in primate primary visual cortex. Science 364:1275-1279.

Gegenfurtner KR, Kiper DC, Fenstemaker SB (1996) Processing of color form and motion in macaque area V2. Vis Neurosci, 13:161-172.

Güçlü U, van Gerven MA (2015) Deep neural networks reveal a gradient in the complexity of neural representations across the ventral stream. J Neurosci 35:10005-10014.

Johnson EN, Hawken MJ, Shapley R (2001) The spatial transformation of color in the primary visual cortex of the macaque monkey. Nat Neurosci 4:409-416.

Kanda Y, Sasaki KS, Ohzawa I, Tamura H (2020) Deleting object selective units in a fully-connected layer of deep convolutional networks improves classification performance. arXiv:2001.07811.

Kandel ER, Koester JD, Mack SH, Siegelbaum SA(Eds.) (2021). Principles of Neural Science, 6e. McGraw Hill.

Kanwisher N, Khosla M, Dobs K (2023) Using artificial neural networks to ask 'why' questions of minds and brains. Trends Neurosci 46:240-254.

Khaligh-Razavi SM, Kriegeskorte N (2014) Deep supervised, but not unsupervised, models may explain IT cortical representation. PLoS Comput Biol 10:e1003915.




Kiefer J, Wolfwitz J (1952) Stochastic estimation of the maximum of a regression function.

    Ann. Math. Stat. 23:462-466.

Komatsu H, Ideura Y, Kaji S, Yamane S (1992) Color selectivity of neurons in the inferior

    temporal cortex of the awake macaque monkey. J Neurosci 12:408-424.

Kriegeskorte N, Mur M, Ruff DA, Kiani R, Bodurka J, Esteky H, Tanaka K, Bandettini PA

    (2008) Matching categorical object representations in inferior temporal cortex of man and

    monkey. Neuron 60:1126-1141.

Krizhevsky A, Sutskever I, Hinton GE (2012) Imagenet classification with deep convolutional

    neural networks. In Advances in Neural Information Processing Systems 27:1097-1105.

Lafer-Sousa R, Conway BR (2013) Parallel, multi-stage processing of colors, faces and shapes

    in macaque inferior temporal cortex. Nat Neurosci 16:1870-1878.

Landisman CE, Ts'o DY (2002) Color processing in macaque striate cortex: relationships to

    ocular dominance, cytochrome oxidase, and orientation. J Neurophysiol 87:3126-3137.

Leavitt ML, Morcos AS (2020) Selectivity considered harmful: evaluating the causal impact of

    class selectivity in DNNs. arXiv:2003.01262.

Leventhal AG, Thompson KG, Liu D, Zhou Y, Ault SJ (1995) Concomitant sensitivity to

    orientation, direction, and color of cells in layers 2, 3, and 4 of monkey striate cortex. J

    Neurosci 15:1808-1818.

Levitt JB, Kiper DC, Movshon JA (1994) Receptive fields and functional architecture of

    macaque V2. J Neurophys, 71:2517-2542.

Livingstone MS, Hubel DH (1984) Anatomy and physiology of a color system in the primate

    visual cortex. J Neurosci 4:309-356.




Livingstone M, Hubel D (1988) Segregation of form, color, movement, and depth: anatomy,

physiology, and perception. Science 240:740-749.

Murphy KP (2012). Machine Learning: A Probabilistic Perspective. Cambridge, MA: The MIT

Press.

Naselaris T, Stansbury DE, Gallant JL (2012) Cortical representation of animate and inanimate

objects in complex natural scenes. J Physiol Paris 106:239-249.

Nassi JJ, Callaway EM (2009) Parallel processing strategies of the primate visual system. Nat

Rev Neurosci 10:360-372.

Nayebi A, Zhuang C, Norcia AM, Kong NCL, Gardner JL, Yamins DLK (2021) Mouse visual

cortex as a limited resource system that self-learns an ecologically-general representation.

bioRxiv, pages 448730.

Olah C, Mordvintsev A, Schubert L (2017) Feature Visualization Distill 00007.

Paszke A, Gross S, Massa F, Lerer A, Bradbury J, Chanan G, et al. (2019). "PyTorch: An

imperative style, high-performance deep learning library," in Proceedings of the 33th

International Conference on Neural Information Processing Systems, Vancouver, 8024-

8035.

Peres R, Soares JGM, Lima B, Fiorani M, Chiorri M, Florentino MM, Gattass R (2019)

Neuronal response properties across cytochrome oxidase stripes in primate V2. J Comp

Neurol 527:651-667.

Peterhans E, von der Heydt R (1993) Functional organization of area V2 in the alert macaque.

European Journal of Neuroscience, 5:509–524.

Shipp S, Zeki S (2002) The functional organization of area V2, I: specialization across stripes




and layers. Vis Neurosci 19:187-210.

Silverman MS, Grosof DH, De Valois RL, Elfar SD (1988) Spatial-frequency organization in

primate striate cortex. Proc Natl Acad Sci USA 86:711-715.

Simonyan K, Zisserman A (2015) Very deep convolutional networks for large-scale image

recognition. arXiv:1409.1556.

Sincich LC, Horton JC (2005) The circuitry of V1 and V2: integration of color, form, and

motion. Annu Rev Neurosci 28:303-326.

Tamura H, Sato H, Katsuyama N, Hata Y, Tsumoto T (1996) Less segregated processing of

visual information in V2 than in V1 of the monkey visual cortex. Eur J Neurosci 8:300-

309.

Tamura H, Tanaka K (2001) Visual response properties of cells in the ventral and dorsal parts of

the macaque inferotemporal cortex. Cereb Cortex 11:384-399.

Tanigawa H, Lu HD, Roe AW (2010) Functional organization for color and orientation in

macaque V4. Nat Neurosci 13:1542-1548.

Tootell RB, Silverman MS, Hamilton SL, Switkes E, De Valois RL (1988) Functional anatomy

of macaque striate cortex. V. Spatial frequency. J Neurosci 8:1610-1624.

Ts'o DY, Gilbert CD (1988) The organization of chromatic and spatial interactions in the

primate striate cortex. J Neurosci 8:1712-1727.

Voss C, Goh G, Cammarata N, Petrov M, Schubert L, Olah C (2021) Branch specialization.

Distill 00024.008.

Wagatsuma N, Hidaka A, Tamura H (2022) Analysis based on neural representation of natural

object surfaces to elucidate the mechanisms of a trained AlexNet model. Front Comput



Neurosci. 2022 16:979258.

Yamins DL, DiCarlo JJ (2016) Using goal-driven deep learning models to understand sensory

cortex. Nat Neurosci 2016 19:356-365.

Yamins DL, Hong H, Cadieu CF, Solomon EA, Seibert D, DiCarlo JJ (2014) Performance-

optimized hierarchical models predict neural responses in higher visual cortex. Proc Natl

Acad Sci USA 111:8619-8624.



## Acknowledgments

We thank Benjamin Knight, MSc., from Edanz (https://jp.edanz.com/ac) for editing a draft of

this manuscript.

## Additional information

**Author contributions**: HT designed the research, conducted the experiments, analyzed the

data, and wrote the paper.

**Competing interests**: The authors declare no competing financial and/or no-financial interests.

**Materials & Correspondence**: Dr. Hiroshi Tamura

Cognitive Neuroscience Group, Graduate School of Frontier Biosciences, Osaka University,

1-4 Yamadaoka, Suita, Osaka 565-0871, Japan

Tel: +81-6-6879-7969; E-mail: tamura.hiroshi.fbs@osaka-u.ac.jp



**Figure Legends**

**Figure 1**. *A*, Architecture of two-streams fully parallel (2SFP) AlexNet. Each stream of 2SFP-AlexNet contains five convolutional layers (conv1−5) with an activation function (ReLU), three pooling layers (Max-pool). Outputs from two streams were combined and fed into fully connected (FC) layers and the output layer for classification. The number of filters in convolutional layers were indicated in parentheses. *B*, Conv1-filters from stream 1 (top) and those from stream 2 (bottom) of three representative instances (left, center, right). For visualization, the minimum and maximum weight values were scaled between 0−255. Single and double asterisks are filters mentioned in the main text. *C*, Comparisons of color index (top), orientation index (middle), and preferred spatial frequency (cycles/filter) of conv1 filters of stream 1 (blue) and stream 2 (orange) for the three instances in B. In the violin plots, a kernel density estimation was provided, and vertical bars show each underlying datapoint. Double asterisks indicate significant differences between two distributions ($p < 0.01$). "ns" indicates non-significant differences.

**Figure 2**. Relationships among color index, orientation index and preferred spatial frequency (SF) of conv1 filters of two-streams fully parallel AlexNet. *A*, Relationships between color index and orientation index (left), between color index and preferred SF (center), and between orientation index and preferred SF (right) in a model instance. This instance is the same as that shown in Figure 1-right. Each circle represents a filter of conv1 of a stream 1 (blue) or stream 2 (orange). Correlation coefficient (*r*) was provided for each panel. *B*, Frequency distributions of correlation coefficient between color index and orientation index (left), color index and



preferred SF (center), and orientation index and preferred SF (right) from 16 model instances.

*C*, Relationships between color index and orientation index (left), between color index and

preferred SF (center), and between orientation index and preferred SF (right) with all of the

conv1 filters from 16 model instances. The number of filters plotted was 1,895. Each circle

represents a filter of conv1. A correlation coefficient (*r*) was provided for each panel.

**Figure 3**. *A*, Comparisons of color index (top), orientation index (middle), and preferred spatial

frequency (SF, cycles/filter) of conv1 filters of stream 1 (s1, blue) and stream 2 (s2, orange) of

two-streams fully parallel AlexNet for the 16 instances. *B*, Comparisons of the median color

index (left) and the median orientation index (center) and the mean preferred SF (right) of

conv1 filters of stream 1 (s1, horizontal axis) and stream 2 (s2, vertical axis) across 16

instances. Closed and open circles are significant and non-significant difference between the

distributions (*p* < 0.01), respectively. The diagonal broken line is the equality line. A correlation

coefficient (*r*) was provided for each panel. Other conventions are as in Figure 1.

**Figure 4**. Comparisons of most effective stimuli (MESs) of conv2−5 filters of stream 1 and

stream 2 of two-streams fully parallel AlexNet. *A-B*, Sixteen examples of MESs each from

conv2−5 of stream 1 (top) and stream 2 (bottom) of a model instance (A) and another instance

(B). *C-D*, Comparisons of color index and preferred SF of conv2−5 filters of stream 1 (blue)

and stream 2 (orange) of A (C) and B (D). Other conventions are as in Figure 1.

**Figure 5**. Relationships between color index and preferred spatial frequency (SF) of most



effective stimuli (MESs) of conv2−5 filters of two-streams fully parallel AlexNet. *A*,

Relationships between color index and preferred SF (cycles/image) of MESs of all of the filters

from all 16 model instances of conv2 (left), conv3 (center-left), conv4 (center-right), and conv5

(right). Each point represents single filter. *B–C*, Comparisons of the median color index (B) and

the median preferred SF (C) of MESs of conv2 (left), conv3 (center-left), conv4 (center-right)

and conv5 (right) filters of stream 1 (horizontal axis) and stream 2 (vertical axis) across 16

instances. *D*, Relationships between the absolute difference between streams in the median

color index of conv1 filters and that of MESs of conv2−5 filters (left). Relationships between

the absolute difference between streams in the mean preferred SF of conv1 filters and that in the

median preferred SF of MESs of conv2−5 filters (right). Each point represents a model instance,

with brown for conv2 filters, pink for conv3 filters, olive for conv4 filters, and cyan for conv5

filters.

**Figure 6**. Comparisons of representational dissimilarity matrix (RDM) between streams of two-

streams fully parallel (2SFP) AlexNet. *A*, RDMs of a model instance calculated with the outputs

from conv1 (left column) and conv5 (right column) filters of stream 1 (top row) and stream 2

(bottom row) to 1,000 stimulus images. Distance between stimulus images were normalized

between 0−1 and plotted in a color scale. *B*, Correlation coefficient between RDMs between

conv layers. Correlation coefficients were color coded. The correlation coefficient between the

same layer (the diagonal element) was the mean across 10 correlation coefficients, each of

which was calculated by randomly dividing filters into two groups. The five numbers on the

plot are correlation coefficients between streams at the same hierarchical level. *C*, Relationship



between absolute difference between streams in the median color index of conv1 filters and

correlation coefficient of RDMs of conv1 filters. Each point corresponds to a single model

instance. ***D***, Changes in correlation coefficients of RDMs along the hierarchy of 2SFP-AlexNet.

Each line corresponds to a single model instance.

**Figure 7**. Effect of deletion of a stream of two-streams fully parallel AlexNet on the

classification accuracy of images and comparisons of properties between inanimate and animate

streams. ***A***, Relationship of proportion correct between original network and network after

deleting stream1 (left) or stream2 (right). Each circle represents one category. The filled blue

circle indicates the most affected category, which shows the largest decrease in proportion

correct, after deletion of stream 1. The filled orange circle indicates the most affected category

after deletion of stream 2. The diagonal broken line is the equality line. ***B***, Comparisons of color

index (top), preferred spatial frequency (SF, middle) and orientation index (bottom) of conv1

filters between the inanimate stream (blue boxes) and animate stream (orange boxes). In the box

plot, the center of each box (black vertical lines) represents the median across the instances,

whereas the top and bottom of the box represent the upper and lower quartiles, respectively. The

attached whiskers connect the most extreme values within 150% of the interquartile range from

the end of each box. ***C***, Comparisons of color index (top) and preferred SF (bottom) of the most

effective stimulus image of conv2−5 filters between the inanimate stream (blue boxes) and

animate stream (orange boxes).

**Figure 8**. ***A***, Visualization of weight of conv1 filters (n = 64) from stream 1 (top) and stream 2



(bottom) of the two-streams fully parallel VGG11 instance. ***B***, Visualization of weight of conv1 filters (n = 64) from stream 1 (top), stream 2 (middle) and stream 3 (bottom) of the three-streams fully parallel AlexNet instance.

**Figure 9**. Schematic summary of the present study. Filters in a stream of two-streams fully parallel AlexNet are orientation selective (conv1) and non-color selective and prefer higher spatial frequency (SF), and the stream contributing to classification of animate images. Filters in the other stream are color selective and prefer lower SF, and the stream contributing to classification of inanimate images.



**Table 1: Model instances**

| Instances | Architecture | Number of streams | Batch size | Initial learning rate |
|-----------|--------------|-------------------|------------|-----------------------|
| a31 | 2SFP-AlexNet | 2 | 128 | 0.01 |
| b02 | 2SFP-AlexNet | 2 | 128 | 0.01 |
| b16 | 2SFP-AlexNet | 2 | 128 | 0.01 |
| c30 | 2SFP-AlexNet | 2 | 128 | 0.01 |
| 101 | 2SFP-AlexNet | 2 | 128 | 0.01 |
| b25 | 2SFP-AlexNet | 2 | 16 | 0.01 |
| b18 | 2SFP-AlexNet | 2 | 32 | 0.01 |
| b30 | 2SFP-AlexNet | 2 | 32 | 0.01 |
| c12 | 2SFP-AlexNet | 2 | 32 | 0.01 |
| c19 | 2SFP-AlexNet | 2 | 32 | 0.01 |
| c21 | 2SFP-AlexNet | 2 | 32 | 0.01 |
| b28 | 2SFP-AlexNet | 2 | 512 | 0.01 |
| c23 | 2SFP-AlexNet | 2 | 128 | 0.02 |
| c27 | 2SFP-AlexNet | 2 | 128 | 0.02 |
| c28 | 2SFP-AlexNet | 2 | 128 | 0.005 |
| 105 | 2SFP-AlexNet | 2 | 128 | 0.005 |
| 405 | 2SFP-VGG11 | 2 | 32 | 0.01 |
| b21 | 3SFP-AlexNet | 3 | 32 | 0.01 |



**Table 2: Effects of deletion of a stream of two-streams fully parallel AlexNet**

| Instances | Deletion of stream 1 | Deletion of stream 2 |
|---|---|---|
| a31 | zebra | rapeseed |
| b02 | giant panda | rapeseed |
| b16 | zebra | rapeseed |
| c30 | zebra | rapeseed |
| 101 | rapeseed | zebra |
| b25 | rapeseed | zebra |
| b18 | zebra | rapeseed |
| b30 | porcupine | European fire salamander |
| c12 | rapeseed | zebra |
| c19 | dugong | rapeseed |
| c21 | zebra | rapeseed |
| b28 | maypole | zebra |
| c23 | rapeseed | zebra |
| c27 | rapeseed | zebra |
| c28 | rapeseed | zebra |
| 105 | zebra | ambulance |

**A**

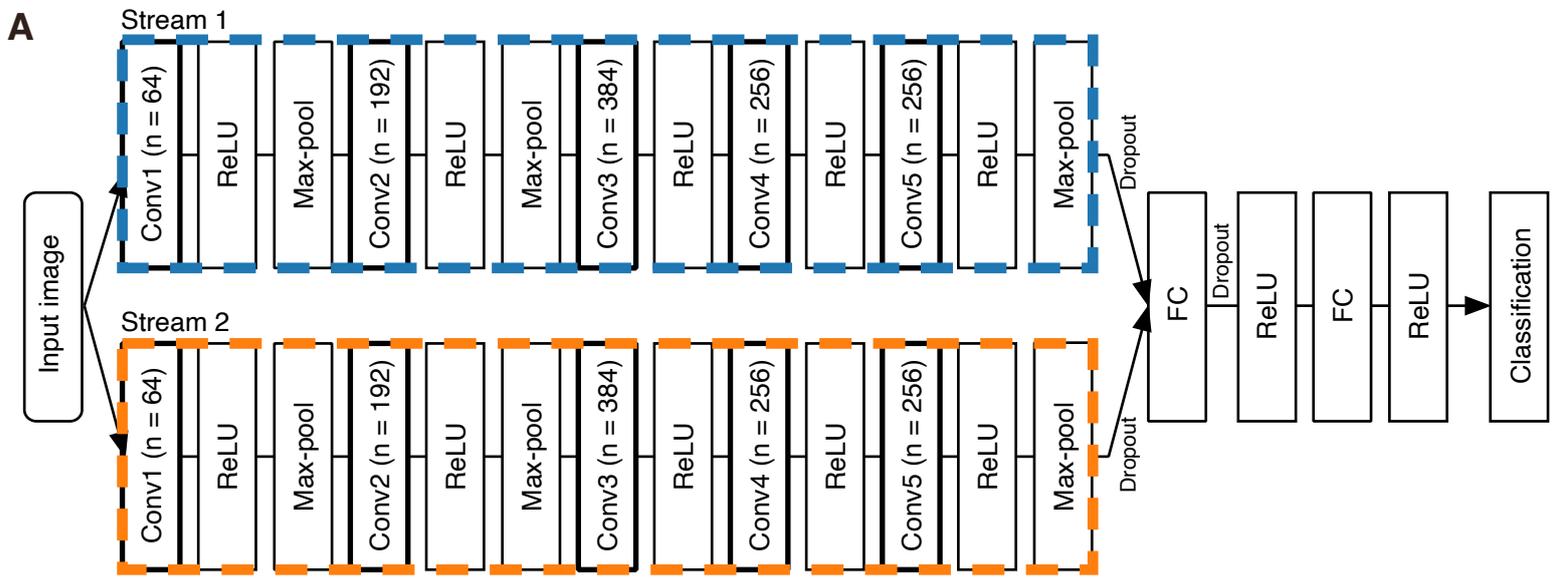

**B**

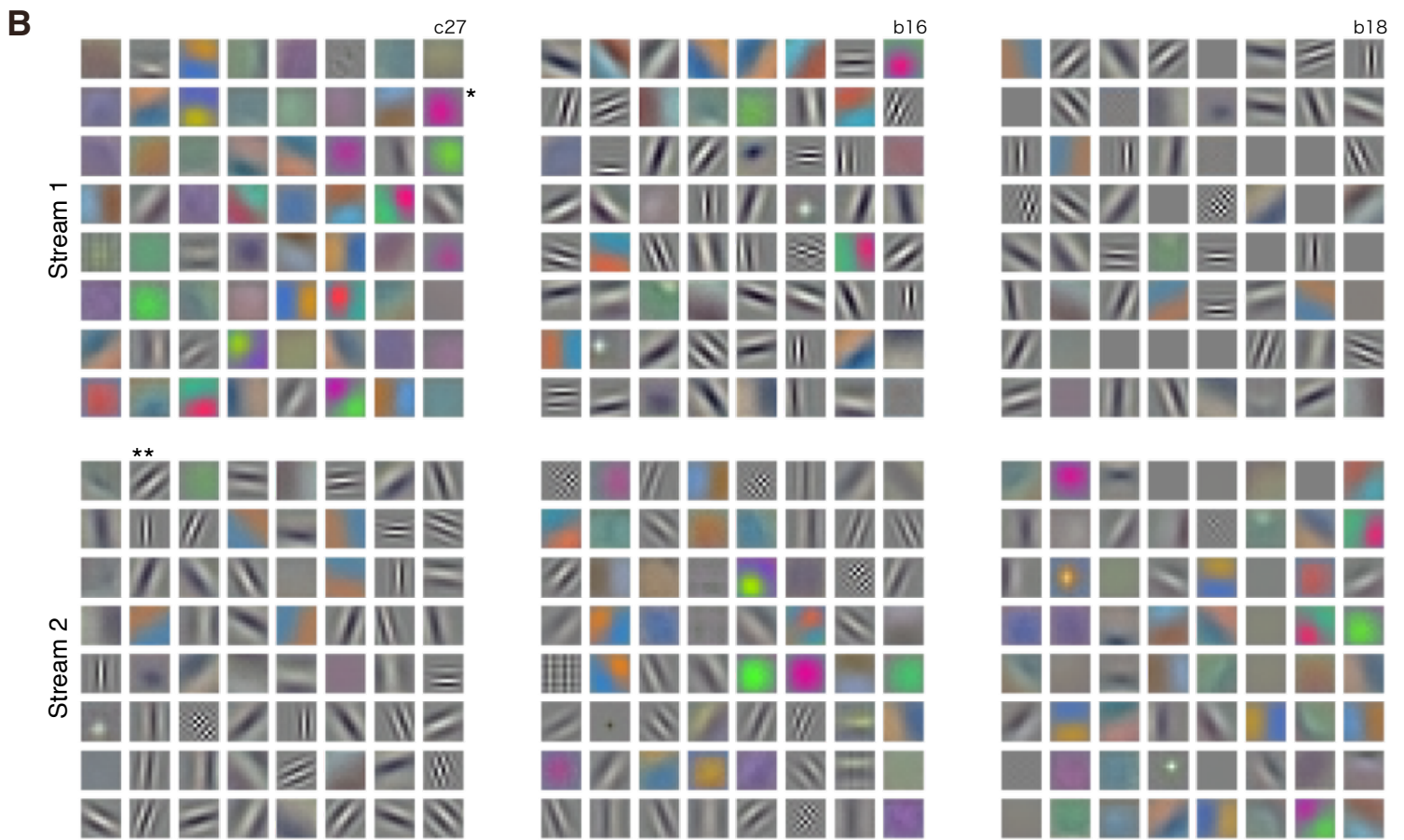

**C**

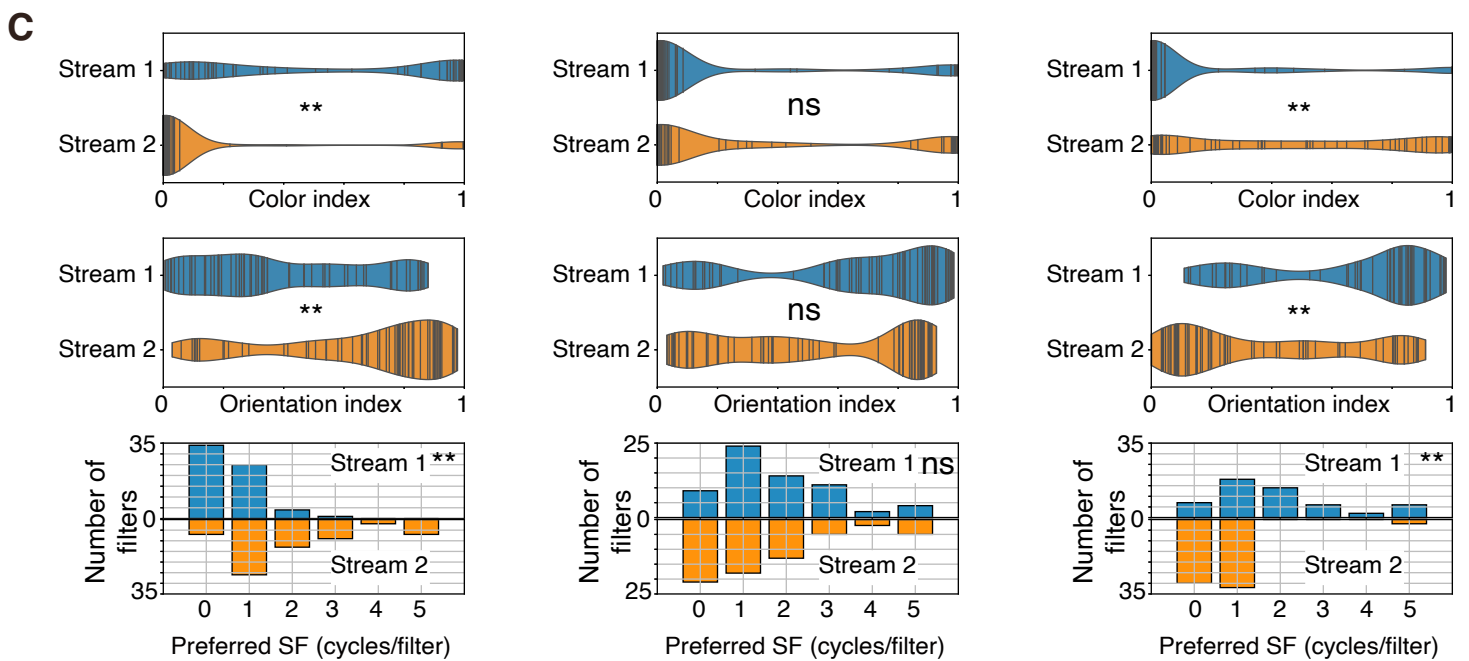

Fig. 1

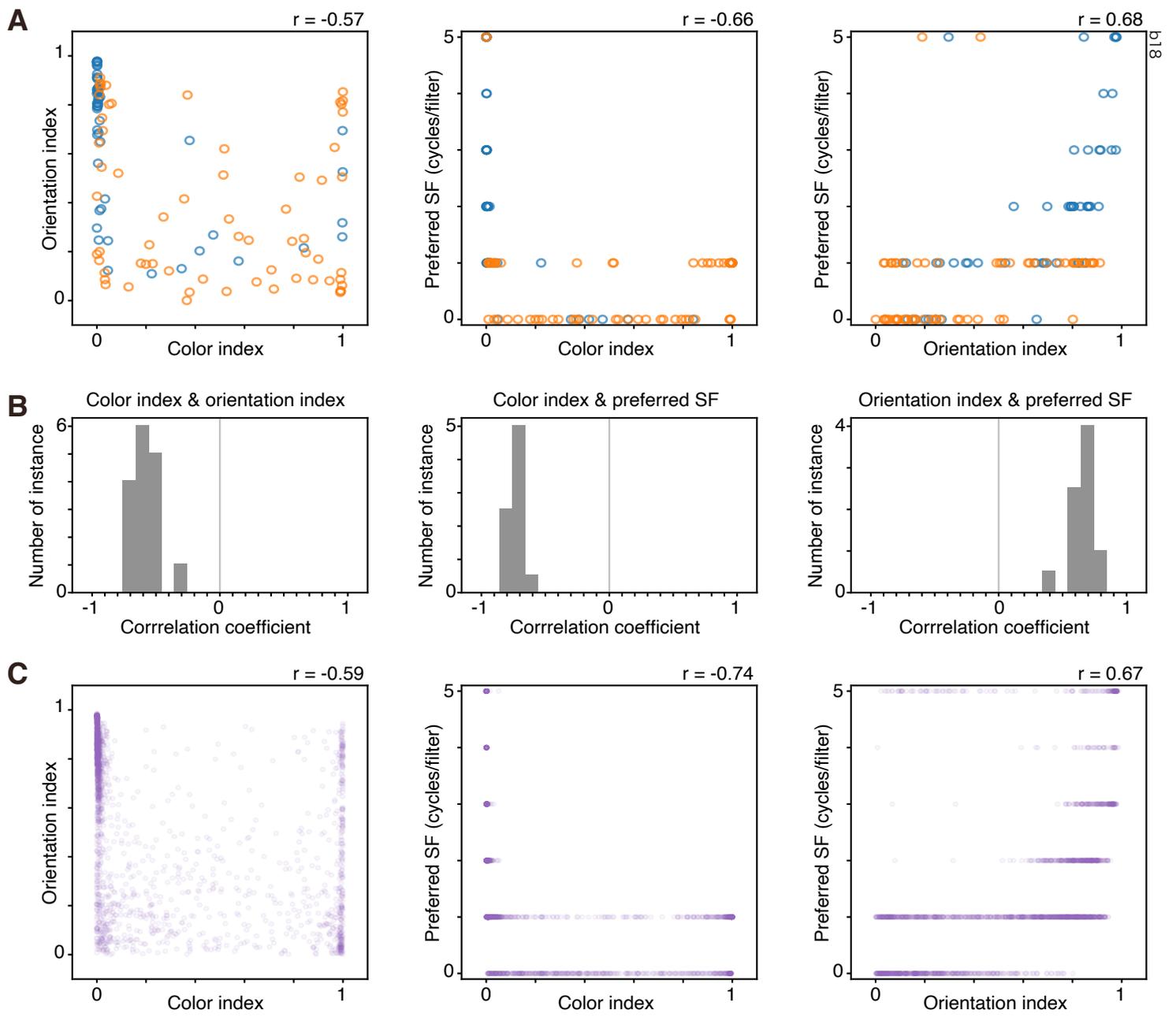

Fig. 2

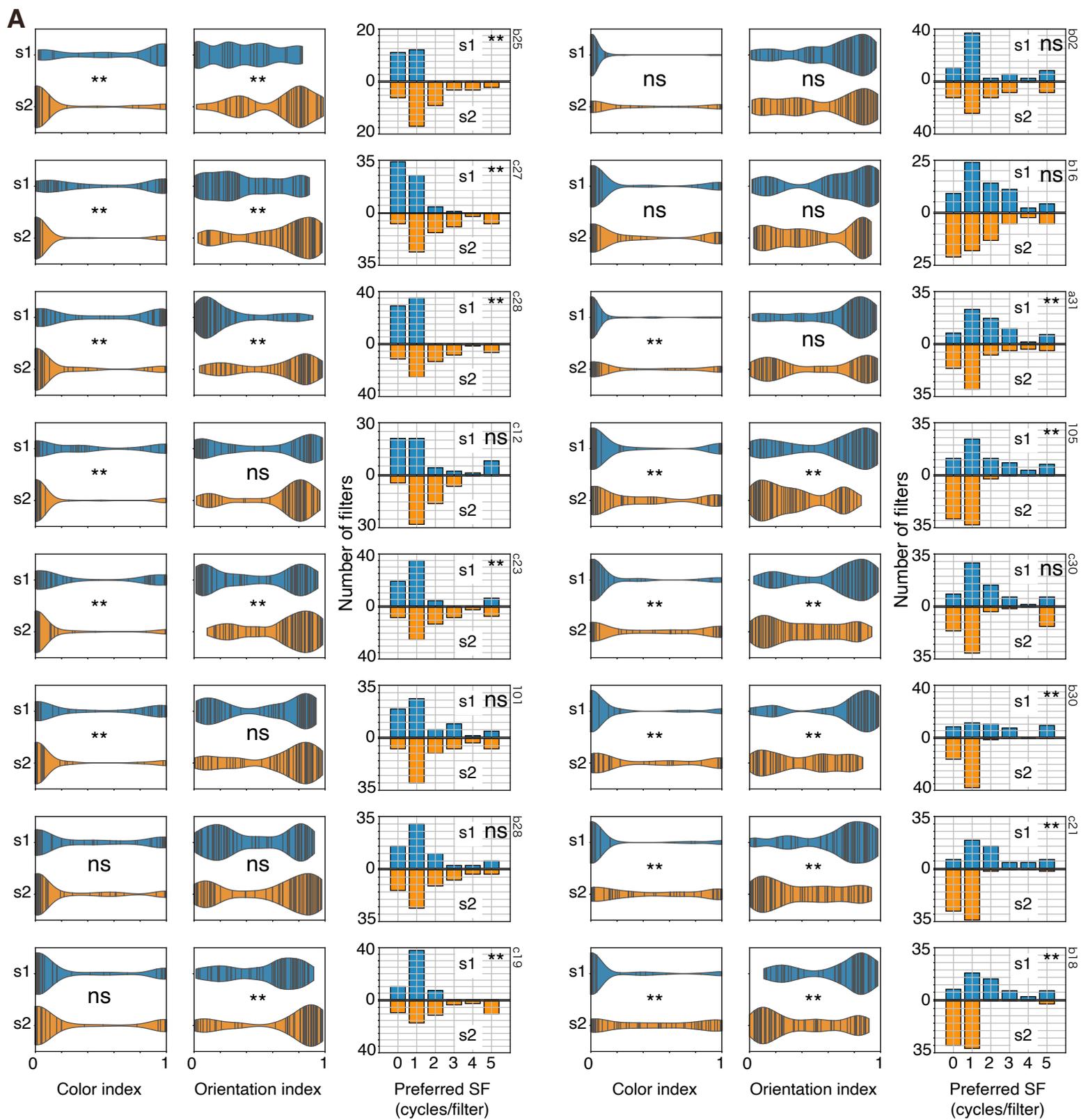

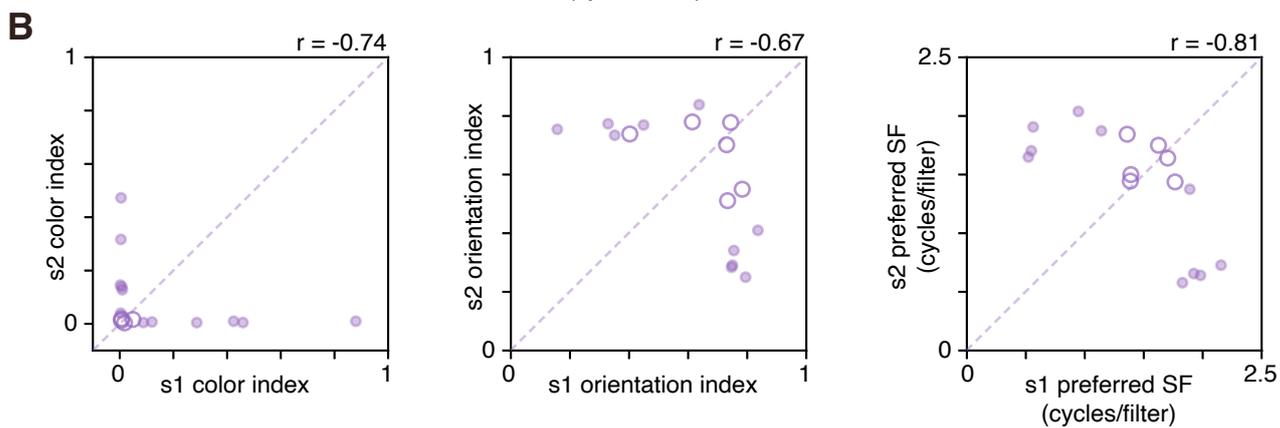

Fig. 3

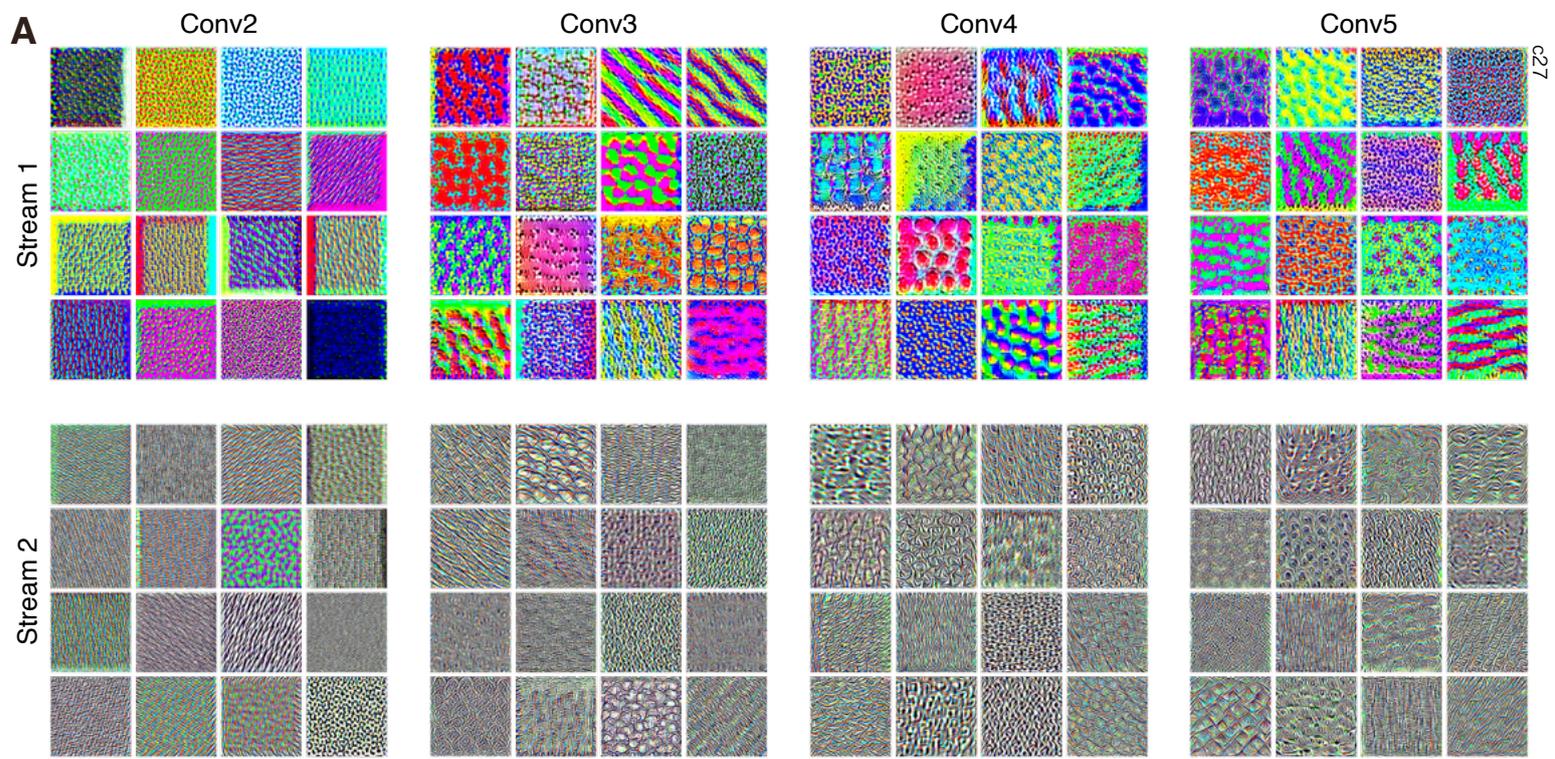

**A**

Conv2    Conv3    Conv4    Conv5

c27

Stream 1

Stream 2

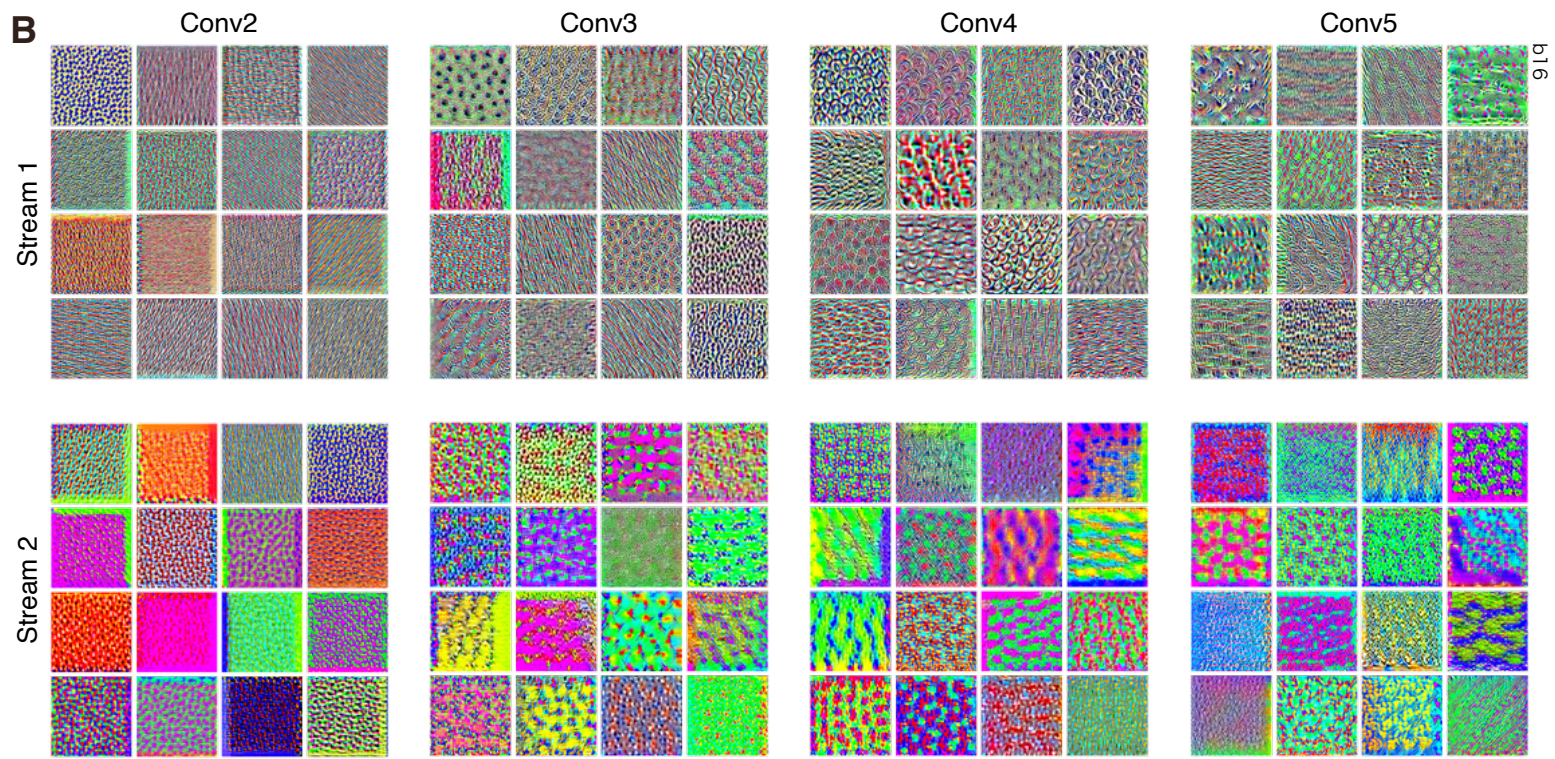

**B**

Conv2    Conv3    Conv4    Conv5

b16

Stream 1

Stream 2

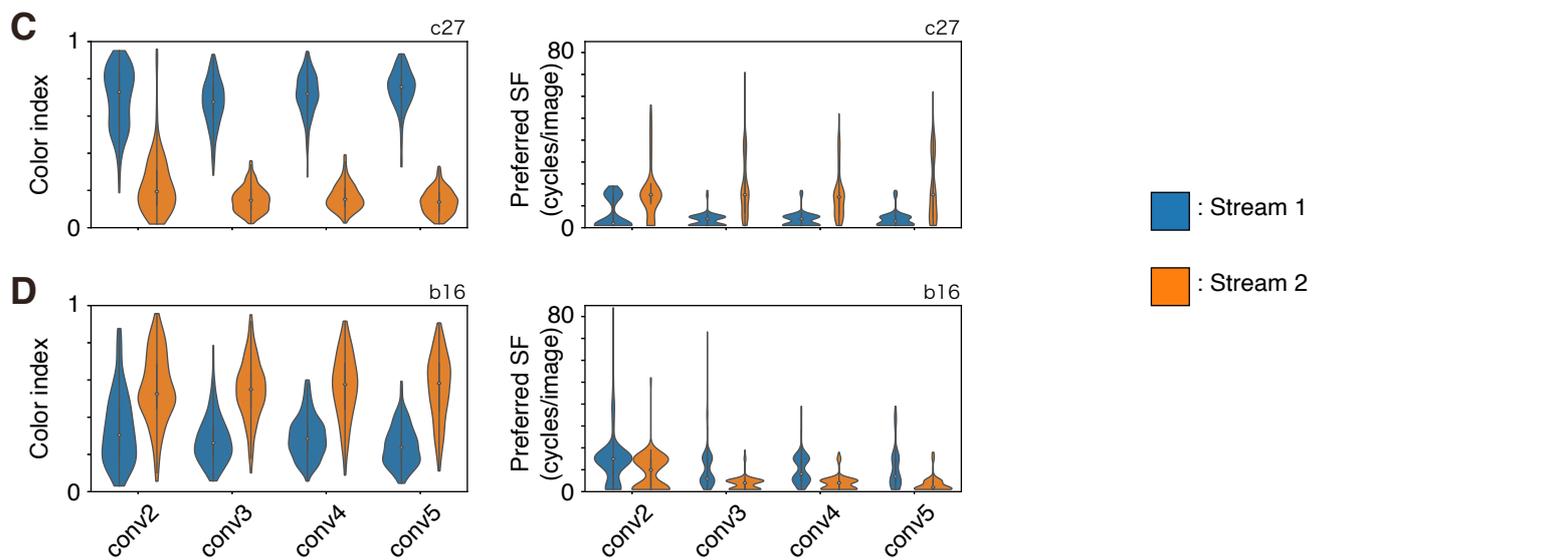

**C**

c27

Color index

c27

Preferred SF
(cycles/image)



0

80

0

**D**

b16

Color index

b16

Preferred SF
(cycles/image)



0

80

0

conv2  conv3  conv4  conv5    conv2  conv3  conv4  conv5

: Stream 1

: Stream 2

Fig. 4

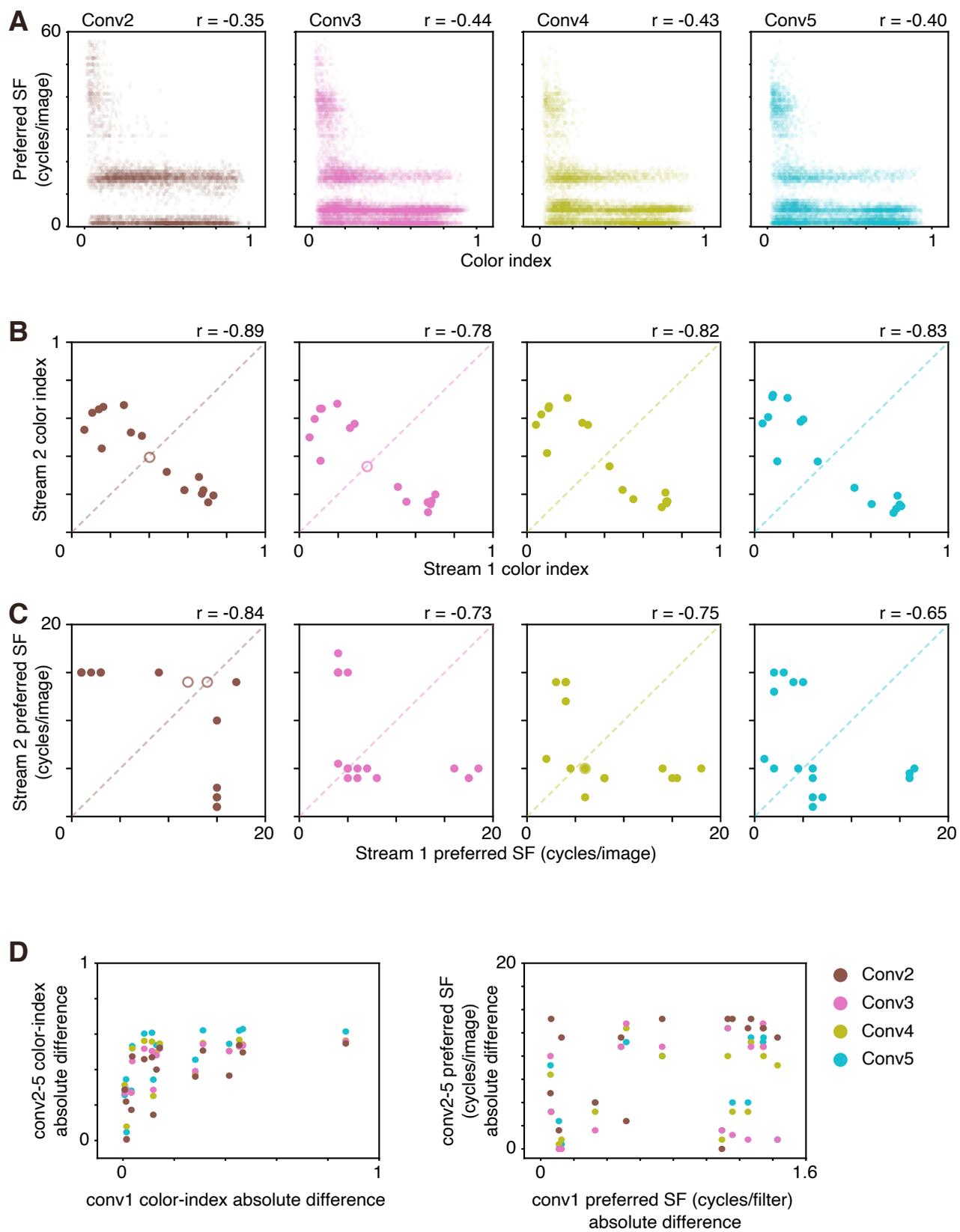



**A**

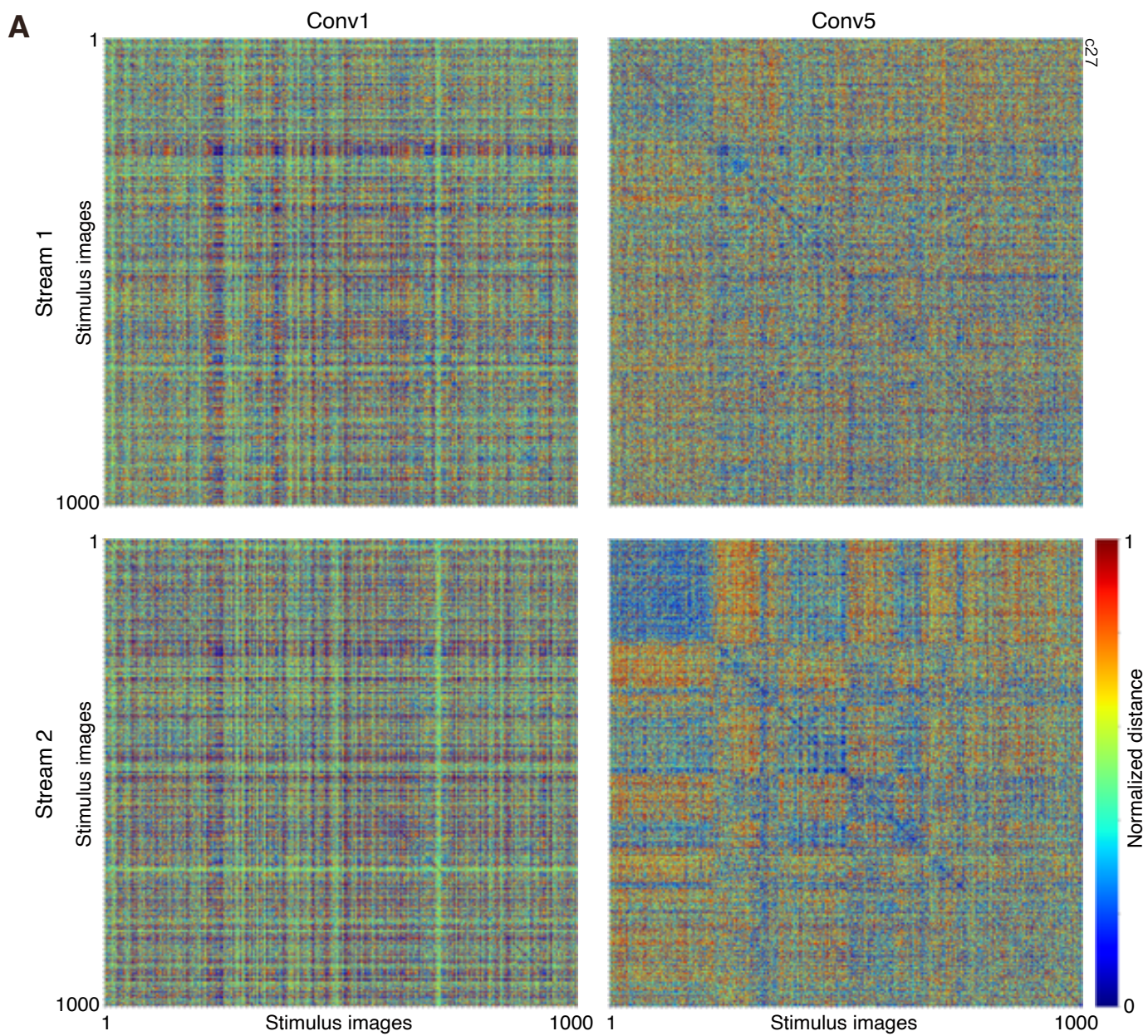

**B**

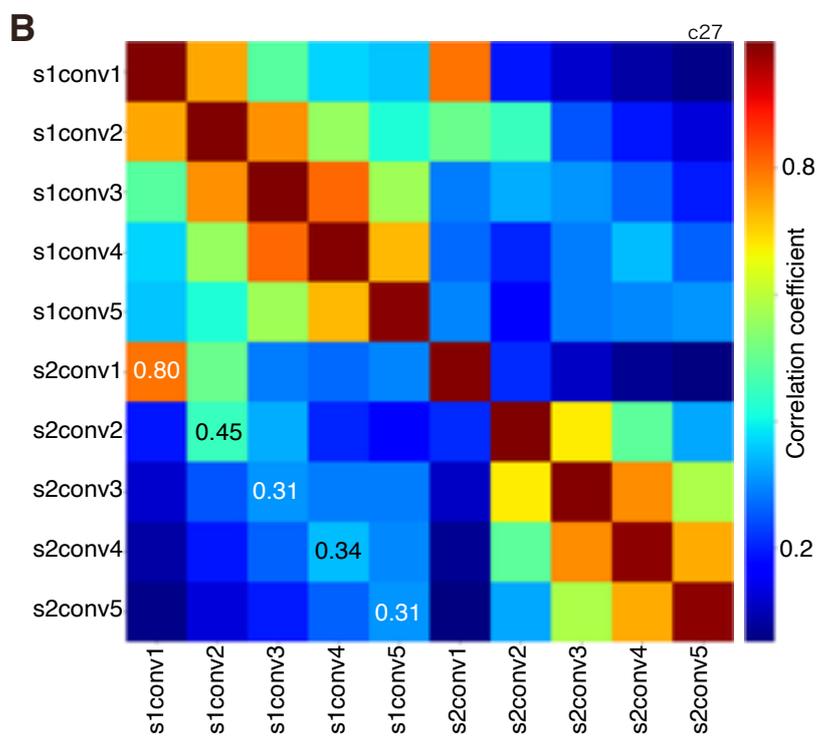

**C**

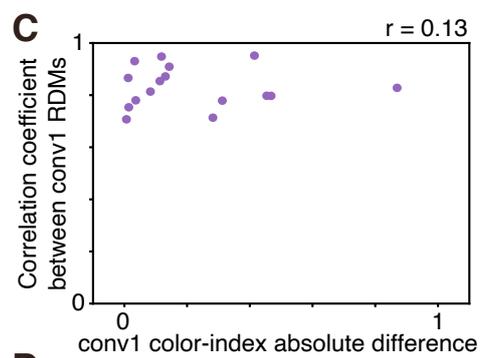

**D**

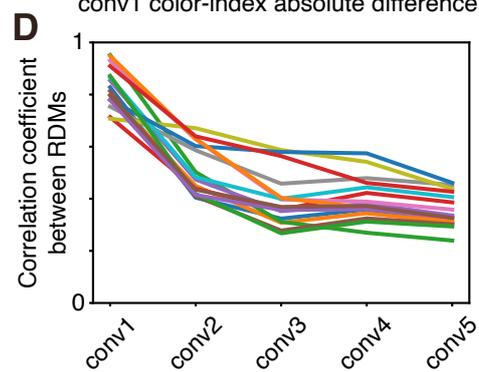

Fig. 6

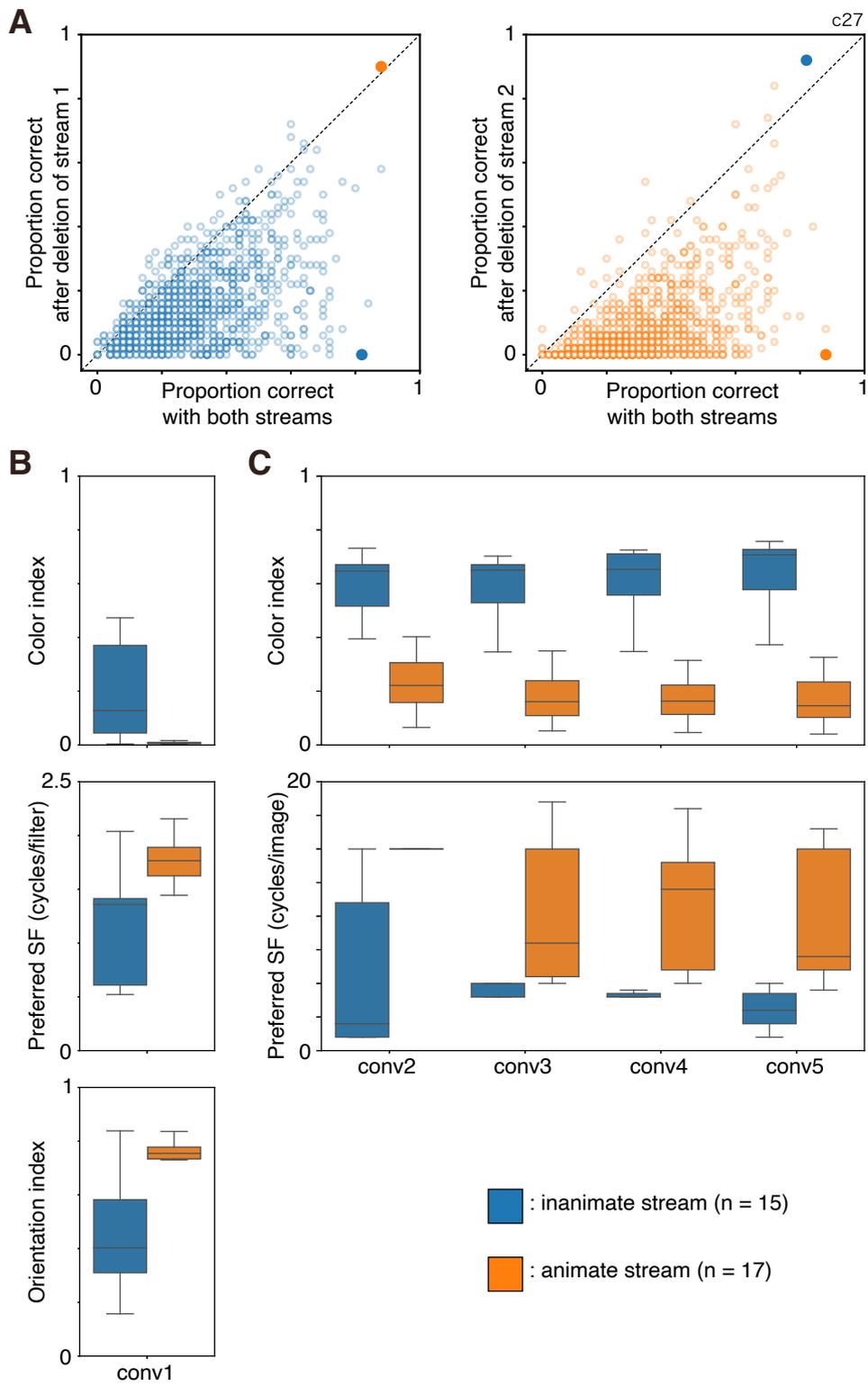



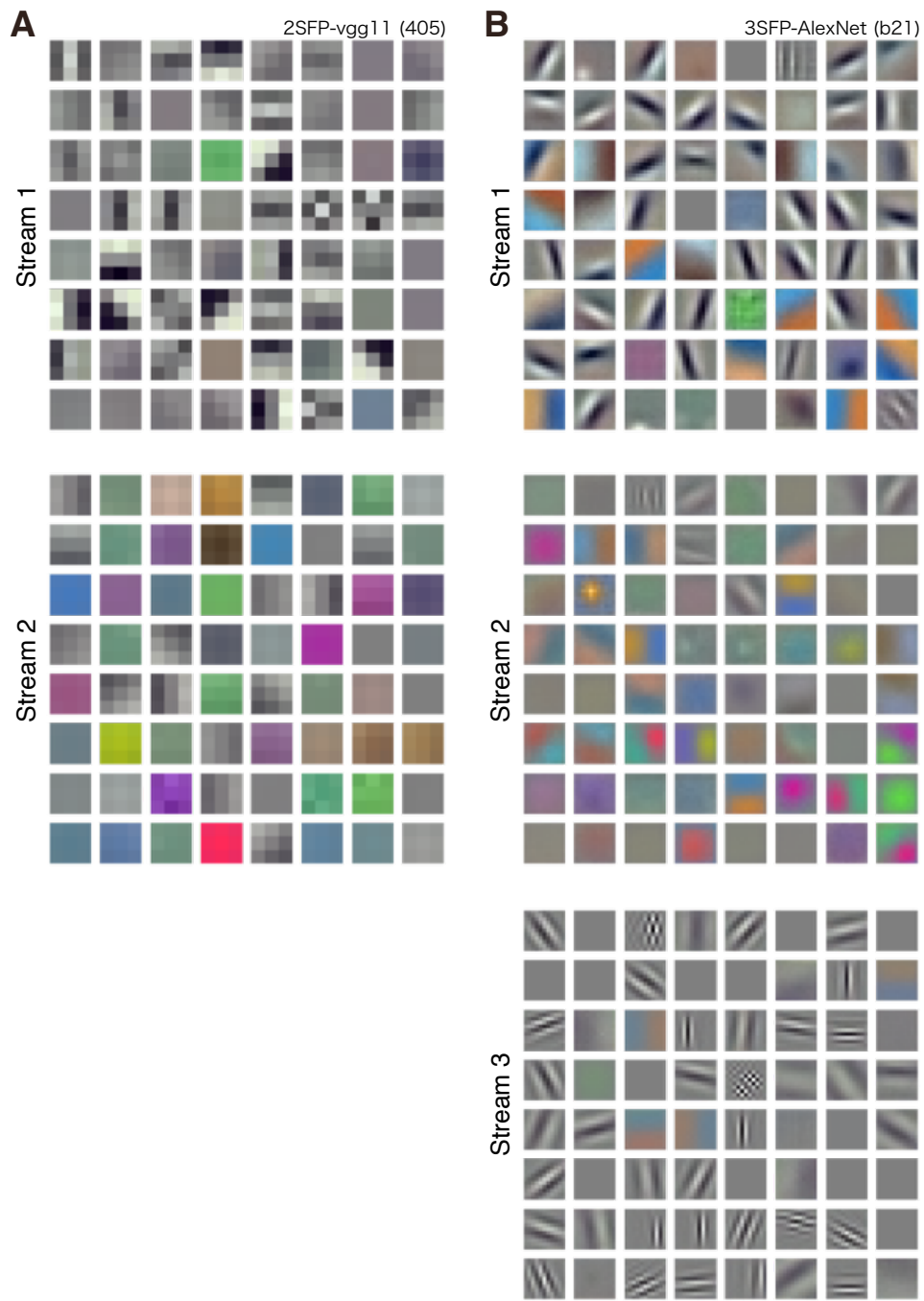

**A** 2SFP-vgg11 (405)

Stream 1

Stream 2

**B** 3SFP-AlexNet (b21)

Stream 1

Stream 2

Stream 3

Fig. 8

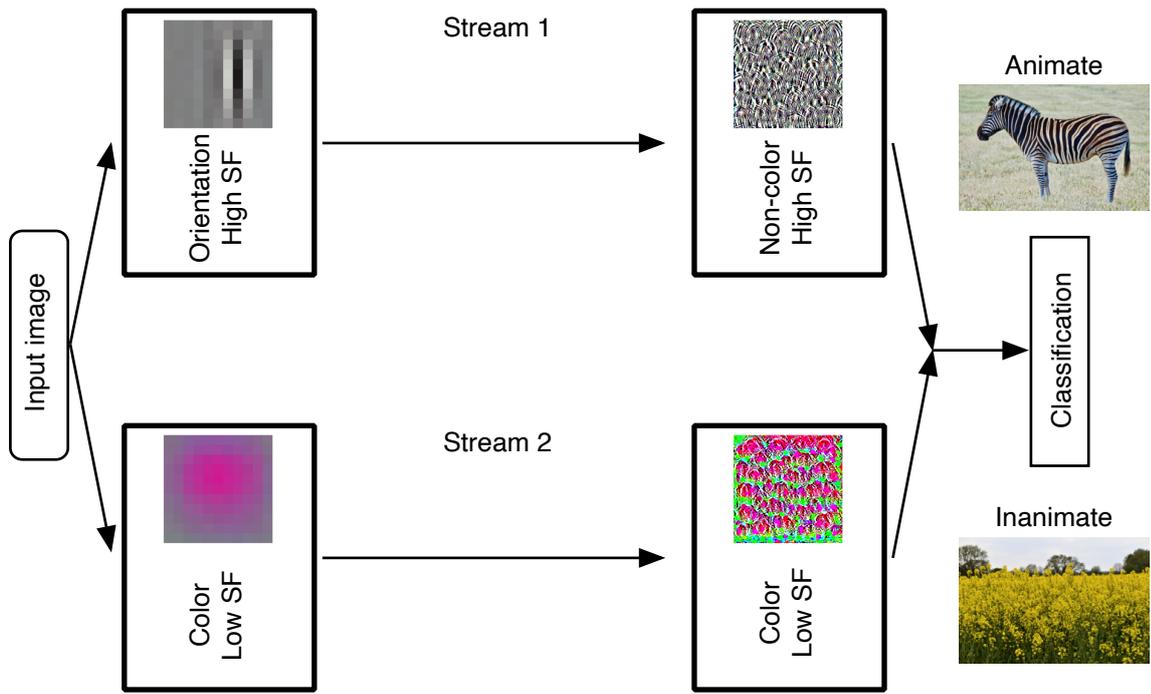

Stream 1

Stream 2

Orientation High SF

Color Low SF

Input image

Non-color High SF

Color Low SF

Animate

Inanimate

Classification